\documentstyle[12pt]{article}
\begin{document}
{\bf
\title{Towards a theory of thin self-gravitating crossing shells}}
\author{Berezin~V.A.  Smirnov~A.L. \\ Institute for Nuclear Research of
Russian Academy of Sciences \\
berezin@ms2.inr.ac.ru \\ smirnov@ms2.inr.ac.ru}
\maketitle
\begin{abstract}
In the case of crossing thin dust shells the momentum conservation law is
found. For two crossing isotropic shells it coincides with the 't~Hooft-Dray
formula. The system of one isotropic and one time-like shell is
considered. In this case we found a very simple formula which relate 
velocities of dust shell before and after crossing.  
\end{abstract}
\newpage
\section{Introduction.}
\label{sec:intro}
General Relativity is a nonlinear theory of gravitational interactions.
Because of this nonlinearity  the inclusion of sources of gravitational field
with full reaction is
a very difficult task. In many physically interesting cases it is possible 
to use very simple idealized source, namely, a thin self-gravitating shell.
The mathematical theory of thin shells in  General Relativity was elaborated
by W.Israel. It was immediately applied to studying  the gravitational collapse 
\cite{2}, but then it was forgotten for a long time. The interest 
to the theory of thin shells  was arised again when it turns out
that the vacuum phase transitions in the early Universe is well described by
bubbles of new vacuum appearing and expanding in a vacuum with high
symmetry \cite{3}. A numerous papers were devoted to studying the 
dynamics of such bubbles \cite{4}. On this way  qualitatively
new results were obtained. In particular, the constrains on parameters of 
the decaying vacuum were found and the 
two bubble universe model was built \cite{5}, \cite{6}. 

The two bubble universe model appeared quite fruitful. Based on it 
V.A. Rubakov and M.E. Shaposhnikov  \cite{7}
showed that the visible world can be a four-dimensional hypersurface in
the fife-dimensional spacetime and fermion fields can be
localized on it. Recently M.Goberishvily \cite{8} , 
L.Randall and R.Sundrum  \cite{9} showed
that the fifth dimension can be noncompact but on a four-dimensional
hypersurface the gravitational field obeys the Einstein equations. These models
name brane-world models. It turns out that such model provide a new solution to
the hierarchy problem. To do this, it is necessary to consider more than 
one brane. Because of that how the problem of crossing shells arises. 

Apart from cosmological applications the thin shells are the simplest objects 
for building quantum black hole models \cite{11},\cite{12},\cite{13} 
and modelling Hawking's radiation \cite{15}.   
But in this area of physics it is necessary to consider crossing shells for
building reasonable models, too.

Studying crossing shells is important for astrophysical applications.
For example, globular clusters is well described by the
system of spherically-symmetric self-gravitating shells. Inside each
shell stars move on quasi-elliptic orbits but total angular momentum
equals zero \cite{16}.
Throughout the paper we use the system of units in which $c=\hbar=1$

\newpage  
\section{Crossing shells.}
\label{sec:crossh}
\subsection{ The momentum conservation law.}
\label{sec:mom_cons}

Conservation laws play important role in physics. There are two basic
conservation laws in classical mechanics and special relativity.
Namely for the momentum and the energy ones.
Both of them are global in these theories.
But the situation is changed in General Relativity. There are not 
global conservation laws in the general case. Nevertheless global 
conservation laws can exist
for  classes of spaces which possess some special properties like 
asymptotically flatness or  symmetries. 
An example of latter class is
static spacetimes. There is the energy conservation law in such spaces.
The energy conservation law will be found in this paper for a system of
self-gravitating crossing thin shells. 

Because of nonlinearity  Einstein equations  even the study of a system of 
thin shells in general is difficult. It is necessary therefore to make
some simplifying assumptions.

Let's a given n-dimensional spacetime with 
evolving shells. One of the assumptions is the choice of the metric
of the form
\begin{equation}
\label{ds1}
ds^2 = g_{AB}(x^0,x^1) dx^A dx^B - R^2(x^0,x^1) g_{ij} dx^i dx^j
\end{equation} 
indices  $A,B = 0,1$ and indices $i,j = 2,...,n-2$, the metric $g_{AB}$ is
a defined function of coordinates $x^2,...,x^{n-2}$, and $R^2$ is a dilaton
field.
Then every shell can be considered
like a classical particle moving in the 2-dimensional pseudo-euclidean 
spacetime. 

Let us assume  that the shells propagate in a vacuum media 
(the vacuum density on one side of a  shell can be different from the 
vacuum density
on the other side. In this case the Einstein equations for the metric (\ref{ds1})
is fully integrable. Indeed,  (2+(n-2))-decomposition of the metric leads to
corresponding decomposition of the Einstein tensor
\begin{equation}
\label{decomp1}
\begin{array}{c}
 {\bf G}_{AB} = \ ^{(2)} {\bf G}_{AB} -\frac{n-2}{R} R_{|AB} 
-\frac{n-2}{R} g_{AB} g^{CD} R_{|CD} +
\frac{(n-2)(n-3)}{2R^2} g_{AB}g^{CD} R_{|C}R_{|D} + \\ \\
\frac{1}{2R^2}g_{AB} \ ^{(n-2)} {\bf R} = 8 \pi G \varepsilon g_{AB}
\end{array}
\end{equation}
\begin{equation}
\label{decomp2}
\begin{array}{c}
{\bf G}_{ik} = \ ^{(n-2)} {\bf G}_{ik} - (n-1)g_{ik}R g^{AC} R_{|AC} -
\frac{(n-2)(n-3)}{2} g_{ik} g^{AC} R_{|A}R_{|C} + \\ \\
\frac{1}{2} g_{ik} R^2 \ ^{(2)} {\bf R} = - 8 \pi G \varepsilon g_{ik} R^2
\end{array}
\end{equation}
Here,  $\ ^{(l)} {\bf G}$, $ \ ^{(l)} {\bf R}$ are the l-dimensional
Einstein tensor and the scalar curvature respectively. The symbol $|$ means a 
covariant derivative with respect to $g_{AB}$. Since the vacuum energy
is constant  $\varepsilon$=const and  $ \ ^{(2)} {\bf G}_{AB}\equiv  0$  
then the metric  $g_{ik}$ is a metric of space with constant unit curvature
i. e. the sphere , the  hyperboloid or the flat space. Thus, for the curvature
we have 
$ \ ^{(n-2)} {\bf R} = k(n-2)(n-3)$, $k= \pm 1,0$ . It is easy to see that
\begin{equation}
\label{cov}
 R_{|AB} = \frac{1}{2} \frac{d \Delta}{d R} g_{AB},
\end{equation}
and the function $\Delta = g^{CD}R_{|C}R_{|D}$ obeys the following
differential equation    
\begin{equation}
\label{diffeq}
 \frac{d \Delta}{d R} + \frac{n-3}{R}\Delta =
\frac{16 \pi}{n-2} G \varepsilon R - \frac{k(n-3)}{R}
\end{equation}
A solution to this equation is
\begin{equation}
\label{sol}
\Delta = \frac{16 \pi G \varepsilon}{(n-1)(n-2)} R^2 + \frac{C_0}{R^{n-1}} -k
\end{equation}
 
It follows from  (\ref{cov}) that the 2-dimensional part of the metric is
\begin{equation}
\ ^{(2)} ds^2 = F dt^2 - F^{-1} dR^2,
\end{equation}
where  $F = - \Delta$. One can show that the equation (\ref{decomp2}) is
satisfied automatically.

At last, let's assume that shell cross each other at the one point. 
Schematically, this system is shown at Fig.1.

Now, we consider two approaches to obtain conservation laws for
systems of crossing shells. The first approach was proposed in \cite {LMD}.
In this work the system of shells satisfies assumptions described
above.

The basic idea of \cite {LMD} is the following. Let us have the trajectory
of N-th shells is $(x^0(\tau),x^1(\tau))$, where $\tau$ is the proper time on
a shell. There exists a local Lorentz transformation from 
the coordinates $(x^0,x^1)$ to the normal coordinates $(\tau,n)$
at every point of the trajectory. Of course, metrics in neighboring 
regions between shells are different. Nevertheless it is possible
to find a coordinate transformation between regions if one consider a small
region in the 
vicinity of the crossing point. This transformations can be done in two
way. We can go around the point clockwise or counter-clockwise.
The necessary condition of self-consistency is identity of the results the 
counter-clockwise transition and the clock-wise transition. By itself 
this identity is trivial. But if one take into account the shells' equations 
of motion then the trivial identity becomes the nontrivial conservation law.

There is another approach to obtaining the
conservation laws. It was proposed in \cite {N}.
This approach doesn't use the local flatness of 2-dimensional space-time
and have some advantage.

Again, there is a system of  N crossing shells. The metric in each 
space-time 
region between two neighboring shells is given by (\ref{ds1}). Let's rename
coordinates  $x^0:=t$, $x^1:=q$ and rewrite (\ref{ds1}) as follows

\begin{equation}
\label{ds2}
ds_a^2 = A_a dt^2 + 2C_a dt dq + B_a dq^2 -
R_a^2(t_a,q_a) g_{ij} dx^i dx^j
\end{equation}

where the index $а = 1 ,..., 2N$ is numerating regions between shells.
We can use the freedom of coordinate transformation and put the following
two conditions

\begin{equation}
\begin{array}{c}
A_a = -B_a \equiv H_a \\
C_a = 0
\end{array}
\end{equation}

Then the 2-dimensional metric takes the form
\begin{equation}
\ ^{(2)} ds^2 = H_a(d \tilde t_a^2 - d \tilde q_a^2)
\end{equation} 

Now, let's introduce the double-null (isotropic) coordinates
\begin{equation}
\label{dnull}
\begin{array}{c}
U_a = \tilde t_a - \tilde q_a \\
V_a = \tilde t_a + \tilde q_a
\end{array}
\end{equation}
then the 2-dimensional metric becomes
\begin{equation}
\label{ds3}
\ ^{(2)} ds^2 = H_a dU_a dV_a
\end{equation}

Here we assume that on the plane $(q,t)$ the coordinate $\tilde q$ varies 
along the
abscissa and it increase to the right and the coordinate $\tilde t$
varies along the ordinate and it goes from down to up.

Now, the metric (\ref{ds2}) with an account of (\ref{ds3}) has a form
\begin{equation}
\label{bds}
ds^2 = H_a(U_a,V_a) dU_a dV_a - R_a^2(U_a,V_a) g_{ij} dx^i dx^j
\end{equation}
As it mentioned above, the  metric (\ref{bds}) has a jump 
crossing the shells. From the other hand,
the metric itself is continuous everywhere.
Hence, we
have to assume the existence of continuous double-null coordinates $(u,v)$. 
In these      
coordinates the metrics in the  vicinity of crossing point reads as follows
\begin{equation}
\label{cds}
ds^2 = h(u,v) du dv - r^2(u,v)g_{ij} dx^i dx^j
\end{equation} 
Here functions $h(u,v)$, $r(u,v)$ are continuous at the crossing point.  

For every region  $a$ coordinates  $(U_a,V_a)$ are
\begin{equation}
\begin{array}{c}
U_a = U_a(u) \\ V_a = V_a(v)
\end{array}
\end{equation}
Then the metric coefficients in (\ref{bds}) and (\ref{cds}) are related by

\begin{equation}
\label{rel}
\begin{array}{c}
h(u,v) = H_a(U_a(u),V_a(v)) U_a^{ \prime}(u) V_a^{\prime}(v) \\
r(u,v) = R_a(U_a(u),V_a(v))
\end{array}
\end{equation} 
Thus, if we sit on the shell between regions  $a$ and $(a+1)$ then
we have
\begin{equation}
\label{con1}
 H_a(U_a,V_a) U_a^{\prime} V_a^{\prime} |_{shell} =
 H_{a+1}(U_{a+1},V_{a+1}) U_{a+1}^{\prime} V_{a+1}^{\prime} |_{shell}
\end{equation}
\begin{equation}
\label{con2}
R_a(U_a,V_a)|_{shell} = R_{a+1}(U_{a+1},V_{a+1})|_{shell}
\end{equation} 
The equation of the shell in region $a$ can be written as follows 
\begin{equation}
\label{sur1}
\Sigma^+_a (U_a,V_a) = 0
\end{equation}
Analogously, for region  $(a+1)$
\begin{equation}
\label{sur2}
\Sigma^-_{a+1} (U_{a+1},V_{a+1}) = 0
\end{equation}
Let's differentiate  (\ref{con2}) by $u$ along the shell. The result is
\begin{equation}
\label{diff}
\left[ \partial_{U_a} R_a + 
\partial_{V_a} R_a \left(\frac{dV_a}{dU_a}\right)_{\Sigma^+} \right] 
U_a^{\prime} = \left[ \partial_{U_{a+1}} R_{a+1} +
\partial_{V_{a+1}} R_{a+1} \left(\frac{dV_{a+1}}{dU_{a+1}}\right)_{\Sigma^-}
\right] U_{a+1}^{\prime}
\end{equation}
The symbol  $( )_{\Sigma}$ means that we should differentiate the implicit
function $V(U)$. This function is given by (\ref{sur1}), (\ref{sur2}) for 
regions $a$ and $(a+1)$ respectively.

It is worth to note that there is symmetry between $u$, $v$ in above equations.
It means that we can obtain equation analogous (\ref{diff}) using $v$ 
instead $u$. If we consider time-like shells the choice of the coordinate
is not crucial. But it becomes crucial for light-light shells.

Let's return to (\ref{diff}). If one consider this equation and condition
(\ref{con1}) at the point of crossing $(u_c,v_c)$ then it is possible
to obtain relationships between $U_a^{\prime},V_a^{\prime}$ 
and  $U_{a+1}^{\prime},V_{a+1}^{\prime}$
\[ \left( 
\begin{array}{c} 
U_{a+1}^{\prime} \\ 
V_{a+1}^{\prime}  
\end{array} 
\right) = T_{a,a+1}
\left( 
\begin{array}{c}
U_a^{\prime} \\ 
V_a^{\prime}
\end{array} 
\right) \]
$T_{a,a+1}$ is a diagonal matrix of transition. This matrix can be 
calculated for each pair of regions.

Let's make successive transition from the region $1$ to the region $2$
and etc., until the circle will be completed. If there exist coordinates
$(u,v)$ in the vicinity of the crossing point, then the obvious 
self-consistent matrix condition should be satisfied

\begin{equation}
\label{beq}
T_{1,2} ... T_{N-1,N} T_{N,1} = I
\end{equation}
$I$ is the identity matrix.

In the next section we'll show that from (\ref{beq}) the momentum
conservation law can be obtain for  time-like dust thin shells and 
isotropic shells in the case of spherical symmetry.

\subsection{The momentum conservation law for spherically-symmetric 
isotropic shells.}
\label{sec:nullsh}
Let's begin this section with the simple 
case of two isotropic shells in vacuum.
The n-dimensional part of the (\ref{beq}) is the metric on the sphere $S^n$.
Then 
\begin{equation}
\label{dsagain}
ds^2 = F(R) dt^2 - F^{-1}(R) dR^2 - R^2 d\Omega_n^2
\end{equation}
The numerating index $a$ is temporarily omitted.
It is convenient to introduce the ``tortoise coordinate''
\begin{equation}
\label{tortoise}
R^* = \int \frac{dR}{F(R)}
\end{equation}
Thus, the metric (\ref{dsagain}) becomes
\begin{equation}
\label{dstortoise}
ds^2 = F(R) [dt^2 - d(R^*)^2] -  R^2 d\Omega_n^2
\end{equation}
From (\ref{dnull}) it follows that the isotropic coordinates are
\begin{equation}
\label{dnull1}
\begin{array}{c}
U = t- R^* \\
V = t+ R^*
\end{array}
\end{equation}
Since, an isotropic shell propagates either along $U$ or $V$ only then 
the equation of shells reads as follows
\begin{equation}
\label{nullsur1}
\Sigma = V - A 
\end{equation}
for the shell propagating along $U$
 \begin{equation}
\label{nullsur1}
\Sigma = U - B
\end{equation}
for the shell propagates along $V$. Here $A$ and $B$ are some constants. 
In the vicinity of the crossing point, the  shells propagate along 
coordinates $u$ and $v$ respectively.
If one take into account the  equality  
\begin{equation}
\left(\frac{dV}{dU}\right)_{\Sigma} =
- \frac{\partial_U \Sigma}{\partial_V \Sigma}
\end{equation}
and $\partial_U \Sigma^{\pm} = 0$ and $ \partial_U R = - F(R)$
then the left and right hand sides of (\ref{diff}) have a form
\begin{equation}
\label{diffn}
(\partial_U R) U^{\prime} = -F  U^{\prime}
\end{equation}
Now we return to numerating of regions and consider the shell between regions 
1 and 2 before the crossing and the same shell between regions 3 and 4 after 
the crossing. Then from (\ref{diffn}) it follows that a transition from
coordinates of region 1 to coordinates of 2 is
\[
 U_2^{\prime} = \frac{F_1}{F_2} U_1^{\prime}
\]
Substituting this expression into (\ref{con1}) we have
\[
 V_1^{\prime} =  V_2^{\prime}
\]
Analogously, for regions 3 and 4 we have
\[
 U_4^{\prime} = \frac{F_3}{F_4} U_3^{\prime}
\]
\[
 V_3^{\prime} =  V_4^{\prime}
\]
If one use the basic identity (\ref{beq}) we can write a conservation law 
for two isotropic shells
\begin{equation}
\label{thooft}
\frac{F_1 F_3}{F_2 F_4} = 1
\end{equation}
This formula was obtained by Dray and 't Hooft  in \cite{17} .  
 
Analogously if the shell propagates along $v$ the transition 
formulae have a form
\[
 V_3^{\prime} = \frac{F_2}{F_3} V_2^{\prime}
\]
\[
 U_2^{\prime} =  U_3^{\prime}
\]
\[
 V_1^{\prime} = \frac{F_4}{F_1} V_4^{\prime}
\]
\[
 U_1^{\prime} =  U_4^{\prime}
\]
Again, we use (\ref{beq}) and obtain (\ref{thooft}).

It is difficult to derive a kind of conservation 
law  from above conclusion. This
obstacle can be overcome if one use formulae from \cite{13}. In this paper
Hamiltonian approach for a self-gravitating thin dust shell was considered.
Authors obtained expression for the canonical momentum conjugate to radius
This expression have a form
\begin{equation}
\label{limit}
\begin{array}{c}
\beta=\exp \left(+\frac{G \hat P_R }{R}\right) = \sqrt{\frac{F_{in}}{F_{out}}} \\ \\ 
\beta=\exp \left(- \frac{G \hat P_R }{R}\right) = \sqrt{\frac{F_{out}}{F_{in}}}
\end{array}
\end{equation}
$ \hat P_R$ is the conjugate momentum, sign  ``-'' in the 
exponent means that the shell is collapsing , sign ``+'' means that
the shell is expanding.
In our case the first shell expands before and after crossing and the second
shell collapses before and after crossing. Thus, we have
for the first shell before crossing
\[
\beta_{b_1} = \exp \left(+\frac{G \hat P_{b_1} }{R}\right) = 
\sqrt{\frac{F_1}{F_2}}
\]
for the second shell before crossing
\[
\beta_{b_2} = \exp \left(-\frac{G \hat P_{b_2} }{R}\right) = 
\sqrt{\frac{F_3}{F_2}}
\]
for the first shell after crossing
\[
\beta_{a_1} = \exp \left(+\frac{G \hat P_{a_1} }{R}\right) = 
\sqrt{\frac{F_4}{F_3}}
\]
for the second shell after crossing
\[
\beta_{a_2} = \exp \left(-\frac{G \hat P_{a_2} }{R}\right) = 
\sqrt{\frac{F_4}{F_1}}
\]
Indices $b$, $a$ means shells before and after crossing.
The momentum conservation law for a system of two isotropic shells reads as
follows
\begin{equation}
\label{lcl}
\beta_{b_1} \beta_{b_2} = \beta_{a_1} \beta_{a_2}
\end{equation}
or 
\[
 \sqrt{\frac{F_1}{F_2}}\sqrt{\frac{F_3}{F_2}} = 
\sqrt{\frac{F_4}{F_3}} \sqrt{\frac{F_4}{F_1}}
\]
i.e. Dray - `tHooft formula.
This conclusion proves that (\ref{thooft}) is the momentum conservation law for
isotropic shells.    

\subsection{Momentum and energy conservation laws for time-like shells.}
\label{sec:timelikes}

In this section we consider a system of crossing time-like shells 
in vacuum. We assume the spherical symmetry of the n-dimensional metric.
The metric has a form (\ref{dsagain}). Again we define ``tortoise
coordinate''.
But in the case under consideration, coordinates $(U,V)$ are defined by more
complicated way, namely
\begin{equation}
\label{adnull}
\begin{array}{c}
U = t- \sigma R^* \\
V = t+ \sigma R^*
\end{array}
\end{equation}
Here $\sigma = +1$, if  $R^*$ increases along the $R$ axis 
(this choice was made above when we define the double-null coordinates),
$\sigma = -1$ if $R^*$ increases to opposite direction of the $R$ axis.

The equation of shells has a form
\begin{equation}
\label{sur3}
\Sigma = R - \hat R(t)
\end{equation}
It is easy to obtain from (\ref{tortoise}) and (\ref{adnull}) the following
relations
\begin{equation}
\label{der}
\begin{array}{c}
\frac{\partial R}{\partial U} = -\frac{\sigma F}{2} \\ \\
\frac{\partial R}{\partial V} = \frac{\sigma F}{2}
\end{array}
\end{equation}
In addition, we have from (\ref{adnull}) 
\begin{equation}
\label{der1}
\begin{array}{c}
\frac{\partial \hat R}{\partial U} = \frac{1}{2}\frac{d\hat R}{dt} \\ \\
\frac{\partial \hat R}{\partial V} = \frac{1}{2}\frac{d\hat R}{dt}
\end{array}
\end{equation}
Substituting (\ref{sur3}) in (\ref{diff}) and using (\ref{der}),
 (\ref{der1}) we obtain
\begin{equation}
\label{inter}
\left( \partial_U R - \partial_V R \frac{ \partial_U \Sigma }
{ \partial_V \Sigma} \right) = \frac{F d\hat R /dt}{F - 
 \sigma d\hat R /dt}
\end{equation}
Let's define a proper time by
\[
d \tau^2 = \left[F - \frac{1}{F} \left(\frac{d\hat R}{dt}\right)^2\right]dt^2
\]
Here, $\rho = \rho(\tau) \equiv \hat R(t)$
It is obvious that
\[
\frac{d\hat R}{dt} = \dot \rho \left(\frac{d\tau}{dt}\right)
\]
Therefore we can rewrite (\ref{inter}) as follows
\begin{equation}
\left( \partial_U R - \partial_V R \frac{ \partial_U \Sigma }
{ \partial_V \Sigma} \right) =
\frac{ F \dot \rho}{\left(\dot \rho^2 +F \right)^{1/2} - \sigma \dot \rho}
\end{equation}
If one takes the latter expression for the shells between regions $a$ and 
$(a+1)$ and remember (\ref{diff}) then
\[
U_{a+1}^{\prime} = \frac{F_a\left( \left(\dot \rho^2 +F_{a+1} \right)^{1/2} - 
\sigma_{a+1} \dot \rho\right)}{F_{a+1}\left(\left(\dot \rho^2 +
F_a \right)^{1/2} - 
\sigma_a \dot \rho \right)} U_a^{\prime}
\]
Let's multiply and divide the above expression by 
$ [\left(\dot \rho^2 +F_{a+1} \right)^{1/2} - \sigma_{a+1} \dot \rho]$ and
$[\left(\dot \rho^2 +F_a \right)^{1/2} - \sigma_a \dot \rho]$ respectively, 
then
\[
U_{a+1}^{\prime} = \frac{\left(\dot \rho^2 +F_a \right)^{1/2} + 
\sigma_a \dot \rho}{\left(\dot \rho^2 +F_{a+1} \right)^{1/2} + 
\sigma_{a+1} \dot \rho} U_a^{\prime}
\]
It follows from (\ref{con1}) that the transition formulae for $V$
have a form
\[
V_{a+1}^{\prime} = \frac{F_a\left( \left(\dot \rho^2 +F_{a+1} \right)^{1/2} + 
\sigma_{a+1} \dot \rho\right)}{F_{a+1}\left(\left(\dot \rho^2 +
F_a \right)^{1/2} + \sigma_a \dot \rho \right)} V_a^{\prime}
\] 
Now, the consistency condition (\ref{beq}) can be written as  
\begin{equation}
\label{multlow}
\prod_{a=1}^{2N} 
\frac{ \left(\dot \rho^2_a +F_{a+1} \right)^{1/2} + 
\sigma_{a+1} \dot \rho_a}{\left(\dot \rho^2_a +F_a \right)^{1/2} + 
\sigma_a \dot \rho_a} = 1
\end{equation}
Now, let us turn  to a system of dust shells. The same
paper \cite{13} gives us an expression for canonical momentum of dust shell
in the case of the spherical symmetry
\[
\beta = \exp \left(\pm\frac{G \hat P_R }{R}\right) =
\frac{\dot \rho +\sigma _{in}\sqrt{\dot\rho^2+F_{in}}}{\dot\rho+\sigma _{out}
\sqrt{\dot \rho^2+F_{out}}}
\] 
Thus, the identity (\ref{multlow}) is the multiplicative momentum 
conservation law i.~e.
\begin{equation}
\label{cons_low}
\prod_{a=1}^{2N} \beta_a \equiv \exp \left(\sum_{a=1}^{2N} \pm 
\frac{G \hat P_R}{\rho_c} \right) =
\prod_{a=1}^{2N} 
\frac{ \left(\dot \rho^2_a +F_{a+1} \right)^{1/2} + 
\sigma_{a+1} \dot \rho_a}{\left(\dot \rho^2_a +F_a \right)^{1/2} + 
\sigma_a \dot \rho_a} = 1 
\end{equation}
Here $\rho_c$ is the radius of the crossing point, and 
$\dot \rho$ is the velocity as a function of the proper time of 
the respective shell.

In addition, we can find the energy conservation law. For simplicity,
we consider a system of two shells. Let's numerate spacetime regions, as
usual. Schwarzschild masses in these region are $m_1$, $m_2$, $m_3$, $m_4$. 
The mass $m_2$ describes the spacetime between shells before  
crossing and the mass $m_4$ describes the spacetime after one.
 The fact that the spacetime is described by two masses instead of
four is based on the Birkhoff uniqueness theorem. If we use the definition of
the total energy a dust shell $\Delta m = m_{out} - m_{in}$, then
\begin{equation}
\label{triv}
\Delta m_{b_1} + \Delta m_{b_2} = \Delta m_{a_1} + \Delta m_{a_2}
\end{equation}
Here,
\[
\Delta m_{b_1} = m_2 - m_1
\]
is the total energy the of the first shell before crossing and etc.
Let's use the equation of motion of a dust shell \cite{6}
\[
\Delta m = \sigma_{in}M \sqrt{\dot\rho^2 + F_{in}} - \frac{GM^2}{2 \rho}, 
 \hspace{5mm}\mbox{где} \hspace{5mm} \Delta m = m_{out} - m_{in}
\]
then the trivial identity (\ref{triv}) become the nontrivial energy
conservation law.

At the end of this section we are going to consider the transition
from a system of time-like shells to a system of light-like shells.

We consider a system of two shells. 
First of all we suppose that parameters $\sigma$ equals $+1$. It means that
the crossing point belong to the $R_+$-region i.e. an observer at infinity
can keep track of shells evolution all time. It is interesting to go to
to the limit
the light speed when shells velocities have equal sign. Substituting
\begin{equation}
\label{asymph1}
(\dot \rho^2 + F)^{1/2} + \dot \rho =
2\dot \rho  + O\left(\frac{1}{\dot \rho}\right)
\end{equation}
into conservation law (\ref{cons_low}) and $ \dot \rho \to \infty $ we
have
\[
1=1
\] 
The physical explanation of such a trivially identity is the following. 
In the case of
time-like shell the shell crossing is possible when the shells move
in the same direction. In other words, one shell can run down another shell.
In the case of isotropic shells this scenario is obviously not possible.
This is why the conservation law reduces to trivial identity.

Thus, the signs of velocities before crossing as well as after crossing
have to be different. 
The momentum conservation law has a form
\begin{equation}
\label{lowtolim}
\begin{array}{c}
\left(\frac{(\dot \rho^2_{b1} + F_2)^{1/2} + \dot \rho_{b1}}
{(\dot \rho^2_{b1} + F_1)^{1/2} + \dot \rho_{b1}}\right)
\left(\frac{(\dot \rho^2_{b2} + F_3)^{1/2} - \dot \rho_{b2}}
{(\dot \rho^2_{b2} + F_2)^{1/2} - \dot \rho_{b2}}\right) \times \\ \\
\times
\left(\frac{(\dot \rho^2_{a1} + F_4)^{1/2} + \dot \rho_{a1}}
{(\dot \rho^2_{a1} + F_3)^{1/2} + \dot \rho_{a1}} \right)
\left(\frac{(\dot \rho^2_{a2} + F_1)^{1/2} - \dot \rho_{a2}}
{(\dot \rho^2_{a2} + F_4)^{1/2} - \dot \rho_{a2}} \right) = 1
\end{array}
\end{equation} 
If we use asymptotic expression
\begin{equation}
\label{asymph2}
(\dot \rho^2 + F)^{1/2} - \dot \rho = 
\frac{F}{2 \dot \rho} + O \left(\frac{1}{\dot \rho^3} \right)
\end{equation}
and (\ref{asymph1}) then (\ref{lowtolim}) is written as follows
$\dot \rho \to \infty$
\begin{equation}
\frac{F_1 F_3}{F_2 F_4} = 1
\end{equation}
i. e. the Dray - 't Hooft formula.   

\subsection{Momentum conservation law for a system of one isotropic 
shell and one time-like shell.}
\label{sec:hybride}

In this section we study a system of isotropic and time-like shell.
There are  several cases of this system. They is shown at 
fig. 2,3,4,5 . Now we have all formulae to find explicit form of (\ref{beq}) 
for each type of system.

First of all let's consider the case in which an isotropic shell
moves from left to right (on the figure). It supposed that shells' inner
region lies always on the left. Now an isotropic shell
propagates along $v$ coordinate then it's better to write transition 
formulae for $V^{\prime}$.
Thus, we have
\[
 V_2^{\prime} = \frac{F_1}{F_2} V_1^{\prime}
\] 
\[
V_3^{\prime} = \frac{F_2\left( \left(\dot \rho_b^2 +F_3 \right)^{1/2} + 
\sigma_3 \dot \rho_b\right)}{F_3\left(\left(\dot \rho_b^2 +F_2 \right)^{1/2} + 
\sigma_2 \dot \rho_b \right)} V_2^{\prime}
\]
\[
V_4^{\prime} = \frac{F_3}{F_4} V_3^{\prime}
\]
\[
V_1^{\prime} = \frac{F_4\left( \left(\dot \rho_a^2 +F_1 \right)^{1/2} + 
\sigma_1 \dot \rho_a\right)}{F_1\left(\left(\dot \rho_a^2 +F_4 \right)^{1/2} + 
\sigma_4 \dot \rho_a \right)} V_4^{\prime}
\]
Momentum conservation law has a form
\begin{equation}
\label{eta_one}
\frac{\left( \left(\dot \rho_b^2 +F_3 \right)^{1/2} + 
\sigma_3 \dot \rho_b\right)\left(\left(\dot \rho_a^2 +F_1 \right)^{1/2} + 
\sigma_1 \dot \rho_a \right)} 
{\left( \left(\dot \rho_b^2 +F_2 \right)^{1/2} + 
\sigma_2 \dot \rho_b\right)\left(\left(\dot \rho_a^2 +F_4 \right)^{1/2} + 
\sigma_4 \dot \rho_a \right)} = 1
\end{equation}

The second case can be named 'reflection', i. e. type of motion
(collapse or expansion) is changed after crossing. In this case we have
\[
V_1^{\prime} = \frac{F_1}{F_2} V_2^{\prime}
\]
\[
V_3^{\prime} = \frac{F_2\left( \left(\dot \rho_b^2 +F_3 \right)^{1/2} + 
\sigma_3 \dot \rho_b\right)}{F_3\left(\left(\dot \rho_b^2 +F_2 \right)^{1/2} + 
\sigma_2 \dot \rho_b \right)} V_2^{\prime}
\] 
\[
V_4^{\prime} = \frac{F_3\left( \left(\dot \rho_a^2 +F_4 \right)^{1/2} + 
\sigma_4 \dot \rho_a\right)}{F_4\left(\left(\dot \rho_a^2 +F_3 \right)^{1/2} + 
\sigma_3 \dot \rho_a \right)} V_3^{\prime}
\]
\[
V_1^{\prime} = V_4^{\prime}
\]
Again, we can write the momentum conservation law as follows
\begin{equation}
\label{eta_one_b}
 \frac{F_1\left( \left(\dot \rho_b^2 +F_3 \right)^{1/2} + 
\sigma_3 \dot \rho_b\right)\left(\left(\dot \rho_a^2 +F_4 \right)^{1/2} + 
\sigma_4 \dot \rho_a \right)}
 {F_4\left( \left(\dot \rho_b^2 +F_2 \right)^{1/2} + 
\sigma_2 \dot \rho_b\right)\left(\left(\dot \rho_a^2 +F_3 \right)^{1/2} + 
\sigma_3 \dot \rho_a \right)} = 1
\end{equation}

Now let's consider cases when an isotropic shell moves from right 
to left.
In the case of crossing we have
\[
U _2^{\prime} = \frac{F_1\left( \left(\dot \rho_b^2 +F_2 \right)^{1/2} - 
\sigma_2 \dot \rho_b\right)}{F_2\left(\left(\dot \rho_b^2 +F_1 \right)^{1/2} - 
\sigma_1 \dot \rho_b \right)} U_1^{\prime}
\]
\[
U _3^{\prime} = \frac{F_2}{F_3} U _2^{\prime}
\]
\[
U _4^{\prime} = \frac{F_3\left(\left(\dot \rho_a^2 +F_4 \right)^{1/2} - 
\sigma_4 \dot \rho_a \right)}{F_4\left(\left(\dot \rho_a^2 +F_3 \right)^{1/2} - 
\sigma_3 \dot \rho_a\right) } U_3^{\prime}
\]
\[
U _1^{\prime} = \frac{F_4}{F_1}U _4^{\prime}
\]
Thus, the momentum conservation law can be written as follows
\begin{equation}
\label{eta_mone}
\frac{\left(\left(\dot \rho_b^2 +F_1 \right)^{1/2} - 
\sigma_1 \dot \rho_b\right)\left(\left(\dot \rho_a^2 +F_3 \right)^{1/2} - 
\sigma_3 \dot \rho_a \right) }
{\left(\left(\dot \rho_b^2 +F_2 \right)^{1/2} - 
\sigma_2 \dot \rho_b\right)\left( \left(\dot \rho_a^2 +F_4 \right)^{1/2} - 
\sigma_4 \dot \rho_a \right) } = 1
\end{equation}
At last, for 'reflection' we have
\[
U _2^{\prime} = \frac{F_1\left(\left(\dot \rho_b^2 +F_2 \right)^{1/2} - 
\sigma_2 \dot \rho_b \right) }{F_2\left(\left(\dot \rho_b^2 +F_1 \right)^{1/2} - 
\sigma_1 \dot \rho_b \right) } U_1^{\prime}
\]
\[
U _3^{\prime} = \frac{F_2}{F_3}U _2^{\prime}
\]
\[
U _4^{\prime} =  U _3^{\prime}
\]
\[
U _1^{\prime} = \frac{F_4 \left(\left(\dot \rho_a^2 +F_1 \right)^{1/2} - 
\sigma_1 \dot \rho_a\right) }{F_1\left(\left(\dot \rho_a^2 +F_4 \right)^{1/2} - 
\sigma_4 \dot \rho_a \right)} U_4^{\prime}
\]
Momentum conservation law has a form
\begin{equation}
\label{eta_mone_b}
\frac{F_3\left( \left(\dot \rho_b^2 +F_1 \right)^{1/2} - 
\sigma_1 \dot \rho_b\right)\left(\left(\dot \rho_a^2 +F_4 \right)^{1/2} - 
\sigma_4 \dot \rho_a \right)}
{F_4\left( \left(\dot \rho_b^2 +F_2 \right)^{1/2} - 
\sigma_2 \dot \rho_b\right)\left(\left(\dot \rho_a^2 +F_1 \right)^{1/2} - 
\sigma_1 \dot \rho_a \right)} = 1
\end{equation}
It is possible to make some formalization of above expressions.
Namely, the first and the second cases turn into the third and fourth
ones respectively if we exchange indices 1 and 3 and simultaneously 
exchange +1 and -1 in front of $\sigma$. Then we can define the parameter 
$\eta$ equal +1 for the first pair of cases and -1 for the second pair 
of cases and it is easy to see that we obtain two generalized formulae 
 \begin{equation}
\label{all_eta}
\frac{\left( \left(\dot \rho_b^2 +F_{2+\eta} \right)^{1/2} + \eta 
\sigma_{2+\eta} \dot \rho_b\right)\left(\left(\dot \rho_a^2 +
F_{2-\eta} \right)^{1/2} + \eta\sigma_{2-\eta} \dot \rho_a \right)} 
{\left( \left(\dot \rho_b^2 +F_2 \right)^{1/2} +\eta 
\sigma_2 \dot \rho_b\right)\left(\left(\dot \rho_a^2 +F_4 \right)^{1/2} +\eta 
\sigma_4 \dot \rho_a \right)} = 1
\end{equation}
\begin{equation}
\label{all_eta_b}
\frac{F_{2-\eta}\left( \left(\dot \rho_b^2 +F_{2+\eta} \right)^{1/2} +\eta 
\sigma_{2+\eta} \dot \rho_b\right)\left(\left(\dot \rho_a^2 +F_4 \right)^{1/2} 
+ \eta \sigma_4 \dot \rho_a \right)}
{F_4\left( \left(\dot \rho_b^2 +F_2 \right)^{1/2} +\eta 
\sigma_2 \dot \rho_b\right)\left(\left(\dot \rho_a^2 +F_{2+\eta} \right)^{1/2} 
+ \eta \sigma_{2+\eta} \dot \rho_a \right)} = 1
\end{equation}
for crossing and 'reflection' respectively.
\subsection{A nice formula.}
\label{sec:formula} 
In this section we obtain a simple expression for velocities before and 
after crossing.
First of all it is necessary to write the system of equations describing
shells. This system consists of momentum conservation law and two equations of
motion (before and after crossing).
If one uses formalized expressions then the system of equations can be written
as
\begin{equation}
\label{sysofeq1c}    
\frac{\left( \left(\dot \rho_b^2 +F_{2+\eta} \right)^{1/2} + \eta 
\sigma_{2+\eta} \dot \rho_b\right)\left(\left(\dot \rho_a^2 +
F_{2-\eta} \right)^{1/2} + \eta\sigma_{2-\eta} \dot \rho_a \right)} 
{\left( \left(\dot \rho_b^2 +F_2 \right)^{1/2} +\eta 
\sigma_2 \dot \rho_b\right)\left(\left(\dot \rho_a^2 +F_4 \right)^{1/2} + \eta 
\sigma_4 \dot \rho_a \right)} = 1
\end{equation}
\begin{equation}
\label{sysofeq2c}
\sigma_2 \left( \dot \rho_b^2 +F_2 \right)^{1/2} -
\sigma_{2+\eta}\left(\dot \rho_b^2 +F_{2+\eta} \right)^{1/2} = 
\eta \frac{G M}{\rho}
\end{equation}
\begin{equation}
\label{sysofeq3c}
\sigma_{2-\eta} \left( \dot \rho_a^2 +F_{2-\eta} \right)^{1/2} -
\sigma_4 \left(\dot \rho_a^2 +F_4 \right)^{1/2} = 
\eta \frac{G M}{\rho}
\end{equation}
We choose  all the parameters $\sigma=1$ 
Then for crossing we have
\begin{equation}
\label{eta1}    
\frac{\left( \left(\dot \rho_b^2 +F_{2+\eta} \right)^{1/2} + \eta 
\dot \rho_b\right)\left(\left(\dot \rho_a^2 +
F_{2-\eta} \right)^{1/2} + \eta \dot \rho_a \right)} 
{\left( \left(\dot \rho_b^2 +F_2 \right)^{1/2} +\eta 
 \dot \rho_b\right)\left(\left(\dot \rho_a^2 +F_4 \right)^{1/2} + \eta 
 \dot \rho_a \right)} = 1
\end{equation}
\begin{equation}
\label{eta12}
\left( \dot \rho_b^2 +F_2 \right)^{1/2} -
\left(\dot \rho_b^2 +F_{2+\eta} \right)^{1/2} = 
\eta \frac{G M}{\rho}
\end{equation}
\begin{equation}
\label{eta13}
\left( \dot \rho_a^2 +F_{2-\eta} \right)^{1/2} - 
\left(\dot \rho_a^2 +F_4 \right)^{1/2} = 
\eta \frac{G M}{\rho}
\end{equation} 

One can use the latter pair of equations to obtain
\[
\left( \dot \rho_a^2 +F_{2-\eta} \right)^{1/2} = 
\frac{\Delta m_a}{M} + \eta \frac{GM}{2\rho}
\]
\[
\left( \dot \rho_a^2 +F_4 \right)^{1/2} = 
\frac{\Delta m_a}{M} - \eta \frac{GM}{2\rho}
\]
\[
\left( \dot \rho_b^2 +F_2 \right)^{1/2} = 
\frac{\Delta m_b}{M} + \eta \frac{GM}{2\rho}
\]
\[
\left( \dot \rho_b^2 +F_{2+\eta} \right)^{1/2} = 
\frac{\Delta m_b}{M} - \eta \frac{GM}{2\rho}
\]
Here  $\Delta m$ means the full energy of shell. Substituting the latter 
expressions into (\ref{eta1}) and making some algebra one can
find very simple expression
\begin{equation}
\label{simple}
\dot \rho_a - \dot\rho_b = \eta \frac{\Delta m_b - \Delta m_a}{M}
\end{equation}

\end{document}